# Quantifying Building Blocks of Life in Planetary Analog Materials: Implications for Prebiotic Chemistry and Biosignature Identification

Xiao'ou Luo[1], Chao He[1,*], Zhengbo Yang[1], Yingjian Wang[1], Ziyao Fang[1], Yu Liu[1], Sai Wang[1], Haixin Li[1]

1. National Key Laboratory of Deep Space Exploration / School of Earth and Space Sciences, University of Science and Technology of China, Hefei 230026, China

*Corresponding authors: Chao He (chaohe23@ustc.edu.cn)

## Abstract

Building blocks of life such as amino acids, nucleobases, and fatty acids are central to prebiotic chemistry and represent key targets in the search for planetary biosignatures. In planetary materials, biomolecules typically occur at trace levels within complex matrices, posing substantial analytical challenges, particularly for quantitative characterization. Here we develop a gas chromatography–tandem mass spectrometry (GC/MS/MS) method that enables robust qualitative and quantitative analysis of 56 prebiotically relevant molecules. The method is applied to a Titan aerosol analog and, for the first time, to a Martian gypsum analog from the Qaidam Basin, revealing diverse inventories of amino acids, nucleobases, and fatty acids in both samples. In the Titan aerosol analog, the first detection of phenylalanine and an extensive inventory of fatty acids, together with elevated nucleobase abundances, offers new insights into atmospheric photochemical synthesis of prebiotic molecules. In the Martian analog sample, amino acids are detectable and exhibit pronounced biotic–abiotic contrasts in abundance patterns relative to those observed in the Titan aerosol analog, whereas fatty acids show more overlapping abiotic and biotic signatures, highlighting the potential of amino acids as robust biosignatures. These results provide quantitative constraints on prebiotic chemical evolution and underscore the utility of GC/MS/MS for biosignature identification in planetary exploration.

## 1. Introduction

The investigation of life's origins and the search for extraterrestrial life are fundamental objectives of planetary exploration. The origin of life on Earth is generally thought to have started from exogenously and endogenously produced prebiotic organic building blocks, including amino acids, nucleobases, and fatty acids. These molecules can be synthesized on Earth through various abiotic pathways such as atmospheric chemistry (Miller 1953; Miller & Urey 1959; Hörst et al. 2012; Ferus et al. 2017; Sebree et al. 2018; Pearce et al. 2024a) and hydrothermal reactions (McCollom et al. 1999; LaRowe & Regnier 2008; Herschy et al. 2014; Zhang et al. 2017; Marlin et al. 2025), or they may be delivered by meteorites (Chyba & Sagan 1992; Burton et al. 2012; Pearce & Pudritz 2015; Naraoka et al. 2023; Glavin et al. 2025). Hypotheses for the origin of life suggest that under favorable conditions, these molecules can polymerize into biological macromolecules (e.g., proteins, RNA, DNA), eventually giving rise to primitive forms of life. However, finding evidence for such prebiotic processes on early Earth is challenging due to the lack of direct geological records. An alternative approach is to study similar ongoing processes on other planetary bodies. For example, on Saturn's moon Titan, photochemistry in the upper atmosphere produces complex organic aerosols. Investigating the formation of life's building blocks in such environment can deepen our understanding of prebiotic chemical evolution. Additionally, the identification of these biomolecules in planetary materials can also aid the search for extraterrestrial life. Compared to more complex and unstable biological macromolecules, these smaller compounds are technically easier to detect and have a greater potential for preservation over long periods. Consequently, they may serve as biosignatures, providing evidence of extinct or present life on potentially habitable planets such as Mars.

Therefore, detecting these building blocks of life, regardless of their abiotic or biotic origin, provides valuable insights into the evolution of life and has become a key focus in planetary exploration missions. Their analysis is typically carried out using gas or liquid chromatography coupled with mass spectrometry (e.g., quadrupole, Orbitrap, or time-of-flight mass spectrometry). Among these techniques, gas chromatography–mass spectrometry (GC/MS) is widely employed owing to its high separation efficiency and access to extensive reference mass spectral libraries. Several planetary missions have carried GC/MS instruments for in-situ analysis, aiming to detect biomolecules on Mars, Comet 67P, and Titan. Although biomolecules have not been confirmed on these bodies, a variety of other organic molecules have been detected (Niemann et al. 2005; Freissinet et al. 2015; Altwegg et al. 2017; Freissinet et al. 2025). In laboratories, where advanced instruments and optimized sample preparation procedures are available, numerous biomolecules have been successfully identified in planetary materials or their analogs using GC/MS. Meteorites represent the most extensively studied planetary materials, in which amino acids (Kvenvolden et al. 1970; Engel & Nagy 1982), nucleobases (Martins et al. 2008), carboxylic acids (Yuen et al. 1984; Huang et al. 2005), and carbohydrates (Cooper et al. 2001; Furukawa et al. 2019) have been identified and quantified. Some of these molecules, including five nucleobases (cytosine, adenine, thymine, guanine, and uracil) and two amino acids (alanine and glycine), have also been detected in planetary analog materials such as laboratory-produced Titan's organic aerosol analogs (Hörst et al. 2012). However, the sensitivity of GC/MS is often limited by background noise. The gas chromatography–tandem mass spectrometry (GC/MS/MS) technique, employing a triple-quadrupole configuration, can improve detection limits by up to two orders of magnitude when operated in selected reaction monitoring (SRM) mode. In this mode, selected parent ions undergo controlled fragmentation to generate diagnostic daughter ions, which are selectively monitored to suppress background interference. Therefore, this technique provides more structural information than traditional GC/MS and enables more reliable identification of trace species in complex samples.

GC/MS/MS has been widely applied to the analysis of biomolecules in both planetary materials—such as meteorites (Timoumi et al. 2025) and returned asteroid samples (Furukawa et al. 2025; Glavin et al. 2025; Mojarro et al. 2025)—and planetary analog materials including photochemical organic aerosol analogs (Sebree et al. 2018; Pearce et al. 2024a). These studies demonstrate the strong capability of GC/MS/MS for sensitive, reliable detection and quantitative measurement of diverse prebiotic compounds (amino acids, nucleobases, sugars, and other prebiotic species) in complex matrices. Such quantitative information is crucial for understanding prebiotic chemistry and assessing potential biosignatures. First, quantifying the concentrations of these biomolecules can help evaluate the contributions of specific prebiotic sources and determine whether polymerization reactions are feasible under plausible prebiotic conditions. Second, quantitative constraints can also provide important insights into the potential origin of these compounds, as the patterns of their relative abundances can indicate whether they are of abiotic or biotic origin. For example, abiotic processes generally favor the formation of simpler amino acids, whereas more complex amino acids are typically enriched in biotic systems (Dorn et al. 2011). Despite these strengths, quantitative analyses of biomolecular abundances using GC/MS/MS have remained relatively limited in studies of planetary materials and their analogs. Moreover, fatty acids, a key class of prebiotic molecules linked to protocell formation, have not been systematically investigated in such analyses.

To address these gaps, we developed a GC/MS/MS method capable of quantitatively analyzing 56 biomolecules-including amino acids, nucleobases, fatty acids and other prebiotic precursors-using a single, unified protocol. We applied this method to two planetary analog materials, including a Titan aerosol analog sample and a Martian analog sample from the Qaidam Basin, to evaluate its utility for interpreting molecular inventories relevant to prebiotic chemistry and potential biosignatures in planetary environments.

## 2. Methods

### 2.1. Sample Preparation

Titan aerosol analogs were prepared using an electron cyclotron resonance (ECR) device, as described in our previous work (Liu et al. 2023; Yang et al. 2025). In this experiment, a gas mixture of $N_2$, $CH_4$, and CO (94.8%/5%/0.2%, each with a purity of 99.99%) was introduced into a vacuum chamber at a total gas flow rate of 50 standard cubic centimeters per minute (sccm), with the chamber pressure settled at 15 Pa. The CO fraction was intentionally elevated relative to Titan's atmosphere (~10 ppm) to promote the formation of O-bearing biomolecules and facilitate their detection. The sample serves as a representative case rather than to reproduce Titan-specific chemistry. The gas mixture was exposed to an ECR plasma in the chamber to simulate the energy sources driving photochemical reactions in Titan's upper atmosphere. After 3 hours of reaction, we collected ~1 g of solid-phase products and dissolved 20 mg of them in 2 mL of methanol (HPLC grade). The solution was vortexed for 5 minutes, and then centrifuged at 11000 rpm for 10 minutes to obtain the supernatant for subsequent processing.

Martian analog samples were collected from the Qaidam Basin located in the north of Tibetan Plateau in northwest China, one of the world's largest and driest playas with an average elevation of ~2800 m. Since the uplift of the Tibetan Plateau, the area has undergone progressive aridification and irradiation, conditions that closely resemble the Martian surface environment (Anglés & Li 2017; Xiao et al. 2017). Gypsum crystals collected from the Xiaoliangshan Dome in the Dalangtan Area (38°28.85′, 91°21.60′) were used in this study (Li et al. 2024). We first selected a relatively large crystal without visible cracks and cleaned the surface with methanol to minimize surface contamination. Then we crushed it to smaller pieces, collected several interior fragments, and ground them into fine powders. For extraction, 200 mg of the powder was mixed with 500 μL of methanol, ultrasonicated for 15 minutes at room temperature, and then centrifuged at 11000 rpm for 10 minutes to collect the supernatant. The procedure was repeated twice, and the resulting supernatants were combined. All equipment used in the whole procedure was pre-washed with methanol and baked in an oven at 80 °C overnight.

The sample solutions were derivatized before GC/MS/MS analysis to form silylated derivatives with lower boiling points and enhanced thermal stability, which allows them to volatilize in the GC inlet without decomposition. The derivatization procedure is modified from that used in previous studies (Sebree et al. 2018; Pearce et al. 2024a). 200 μL of Titan aerosol analog solution and 1.2 mL of the gypsum extract (volumes adjusted based on biomolecule concentrations estimated from preliminary experiments) were added to GC vials and dried at 50 °C under nitrogen (99.999%) flow for 30 minutes. 200 μL of dichloromethane (DCM, ≥ 99.8%, purchased from Sigma-Aldrich) was added to each vial and then dried again at 50 °C under $N_2$ flow to remove any residual water. For the silylation step, 30 μL of dimethylformamide (DMF, ≥ 99.8%, purchased from Sigma-Aldrich) and 30 μL of N-tert-butyldimethylsilyl-N-methyltrifluoroacetamide (MTBSTFA, ≥ 97%, purchased from Sigma-Aldrich) were added. The vials were sealed after purged by $N_2$ flow for 30 seconds to avoid evaporation of reagents, followed by 30 minutes of reaction at 80 °C. Finally, 300 μL of DCM was added to dissolve derivatized products. Two parallel samples were carried through this procedure, along with two procedural blanks containing pure methanol for contamination check.

### 2.2. GC/MS/MS Analysis

To develop a targeted SRM method for analyzing biomolecules of interest, we first tested a series of standards (Table 1). The purchased or prepared standard solutions included: (1) a physiological amino acid standard (Sigma-Aldrich) containing 27 amino acids and related compounds in 0.1 M HCl; (2) a nucleobase standard solution containing seven nucleobases dissolved in 0.1 M NaOH; (3) two fatty acid standard solutions containing C12–C18 and C19–C26 fatty acids, respectively, dissolved in DCM; and (4) other standalone standard solutions, with glycolic acid, cysteine, asparagine, lysine, glutamine, ornithine, and tryptophan dissolved in 0.1 M HCl, while histidine and guanidine were dissolved in methanol. All solutions were prepared at 2.5 μmol $mL^{-1}$ for each biomolecule at room temperature. 50 μL of each standard solution was derivatized according to the protocol described above, and 1 μL of the final solution was injected into a Thermo Scientific TSQ 9610 GC/MS/MS system equipped with a TG-5SILMS capillary column. Helium (99.9999%) was used as the carrier gas, with a constant flow rate of 1.3 mL/min. The oven temperature program began at 100 °C, ramped up at a rate of 10 °C/min, and reached a final temperature of 270 °C, which was held for 15 minutes. The temperatures of the GC inlet, transfer line, and ion source were set to 260, 280, and 300 °C, respectively.

The method development procedure is illustrated in Figure 1, using nucleobase standards as an example. Initially, the standard solution was analyzed in full-scan mode (Figure 1 (a)) to determine the retention time of each compound. The mass spectrum of each peak was compared with the standard spectra from the NIST database, and two parent ions with both high intensity and large m/z values were selected (m/z=307 and 193), as shown in Figure 1 (b) for hypoxanthine. These parent ions were then fragmented under three different collision energies (10 eV, 20 eV, and 30 eV), and two representative daughter ions (m/z=166 and 84) with high intensity were selected for each parent ion, as shown in Figure 1 (c) for m/z=193. Subsequently, we optimized the collision energy for each characteristic ion pair to achieve maximum signal intensity (Figure 1 (d)). Method development and optimization were conducted using the AutoSRM software. The same procedures were applied to all 56 compounds listed in Table 1 for the method development (see Table A1 for the list of retention times and SRM transitions). Compared to previous studies (Sebree et al. 2018; Pearce et al. 2024a), we increased the number of characteristic ion pairs in our method to ensure more accurate and robust compound confirmation, while maintaining a low level of background noise.

**Table 1.** Standards Used for GC/MS/MS Method Development

| | |
|---|---|
| Amino Acids | Alanine, Glycine, Sarcosine, Valine, Leucine, Isoleucine, Proline, Serine, Threonine, Phenylalanine, Methionine, Cysteine, Aspartic Acid, Hydroxyproline, Glutamic Acid, Asparagine, Lysine, Glutamine, Aminoadipic acid, Ornithine, Histidine, Tyrosine, Tryptophan, Cystine |
| Nucleobases | Uracil, Thymine, Cytosine, Hypoxanthine, Adenine, Xanthine, Guanine |
| Fatty Acids | Dodecanoic Acid (C12:0 [a]), Tridecanoic Acid (C13:0), Myristic Acid (C14:0), Pentadecanoic Acid (C15:0), Palmitic Acid (C16:0), Heptadecanoic Acid (C17:0), Stearic Acid (C18:0), Nondecanoic Acid (C19:0), Arachidic Acid (C20:0), Heneicosanoic Acid (C21:0), Behenic Acid (C22:0), Tricosanoic Acid (C23:0), Tetracosanoic Acid (C24:0), Pentacosanoic Acid (C25:0), Hexacosanoic Acid (C26:0), Palmitoleic Acid (C16:1 n-7), Linoleic Acid (C18:2 n-6), Oleic Acid (C18:1 n-9), α-Linolenic Acid (C18:3 n-3), γ-Linolenic Acid (C18:3 n-6), Isostearic Acid |
| Other Biomolecules | Glycolic Acid, Urea, Guanidine, Cystathionine |

**Note.**

[a] Conventional lipid nomenclature (C:D n-x), where C represents the total number of carbon atoms, D the number of double bonds, and n-x denotes the position of the first double bond counted from the terminal methyl end of the molecule.

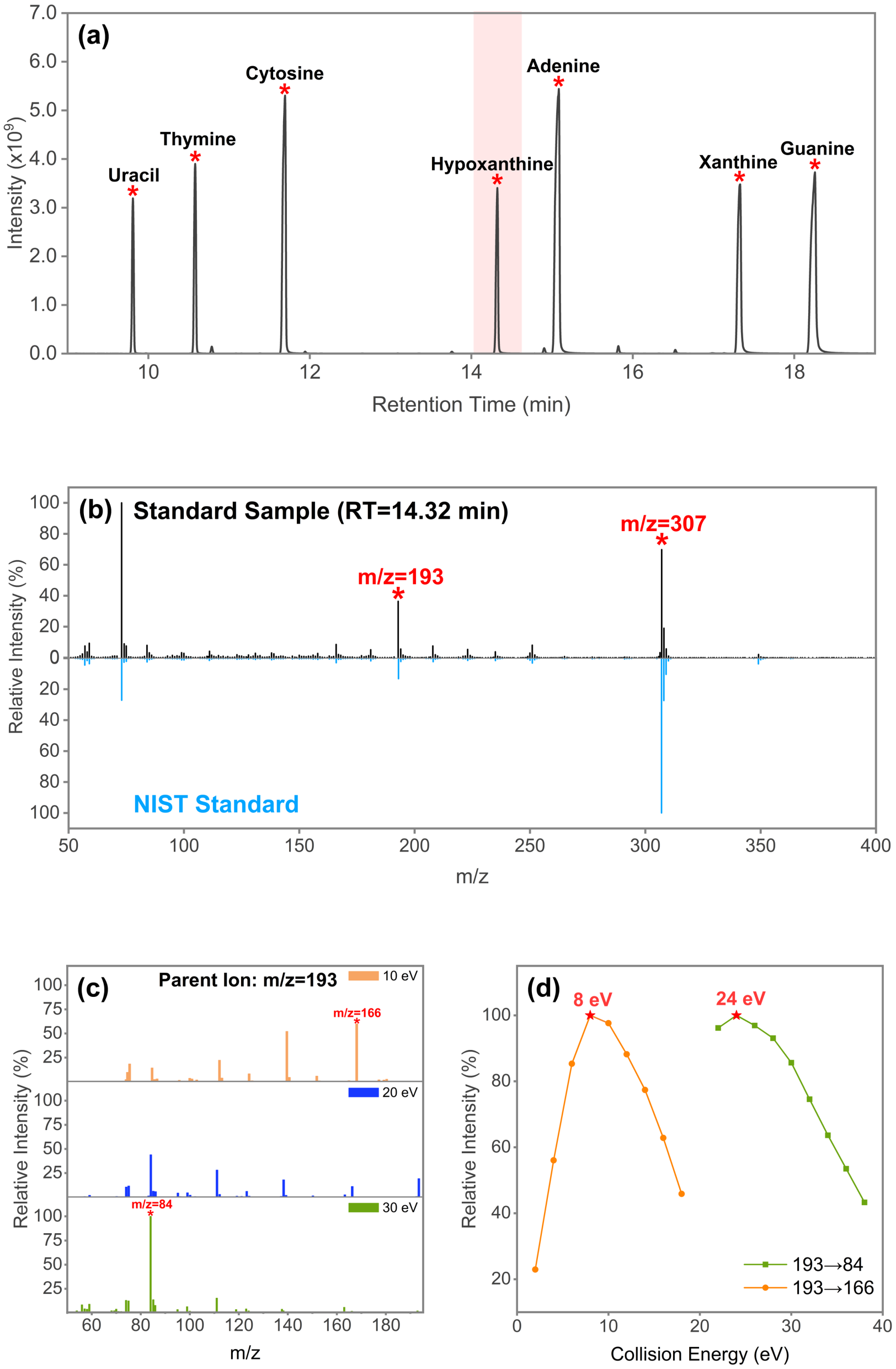

**Figure 1.** Schematic procedure for the SRM method development. (a) Full-scan mode analysis of nucleobase standards. (b) Comparison of the sample mass spectrum at 14.32 min with the NIST standard database spectrum, leading to the selection of m/z 193 and 307 as parent ions for hypoxanthine. (c) Daughter ion spectra derived from the parent ion m/z 193 at collision energies of 10, 20, and 30 eV (signals normalized to the most intense peak, m/z 84 at 30 eV). Ions at m/z 84 and 166 were selected for further analysis. (d) Collision energy optimization for the two selected ion pairs.

The finalized SRM method was then applied to analyze the derivatized sample solutions (Titan aerosol analog and Martian analog samples). An aliquot (1 μL) of each solution was injected into GC/MS/MS for subsequent analysis. Instrumental injection replicates were not performed in order to minimize total sequence time and thereby reduce potential alteration (e.g., degradation, evaporation, or hydrolysis) of derivatized compounds during autosampler storage. Based on repeated injections of calibration standards during method development, the relative standard deviation of peak areas between replicate injections is typically within 2%. Before the sample injection, an instrument blank and two procedural blanks were run on the instrument to establish the baseline level. After each sample injection, a procedural blank run was performed to ensure that the carryover from the previous sample was minimal. The compound confirmation procedure for qualitative analysis is shown in Figure 2. A part of the total ion chromatogram (TIC, Figure 2 (a)) for the Titan aerosol analog sample is used as an example to show two peaks that correspond to the retention times and characteristic ion pairs of hydroxyproline and hypoxanthine, respectively. Although the retention times of the sample peaks differ slightly from those of the standards, a deviation of no more than 0.1 minutes is considered acceptable. Further confirmation was achieved using a processing method developed in the Chromeleon software based on two criteria: the retention time of each characteristic ion peak aligns precisely, and the relative intensities of the four ions match those of the standards. Relative ion intensities were quantitatively evaluated using peak confirmation ratios (PCRs), defined as the ratios of confirmation ion peak areas to that of the quantifier ion. For each compound, the most intense and least interfered ion was selected as the quantifier, and the others as confirmation ions. A Relative Deviation (RD) < 30% between observed and reference PCRs was generally required, relaxed to 50% for low PCR values (< 20%) or near the detection limit; cases with significant interference were assessed individually. In the right half of Figure 2, both the retention time criterion (Figure 2 (c)) and the relative intensity criterion (Figure 2 (e)) are met, which confirms the presence of hypoxanthine in the Titan aerosol analog. In contrast, the left half of Figure 2 shows a discrepancy in the retention times of the four ion pairs (Figure 2 (b)) and significant differences in the relative intensities compared to the standards (Figure 2 (d)), leading to the exclusion of hydroxyproline. This demonstrates the method's ability to effectively exclude false-positive signals, which makes it suitable for analyzing complex samples.

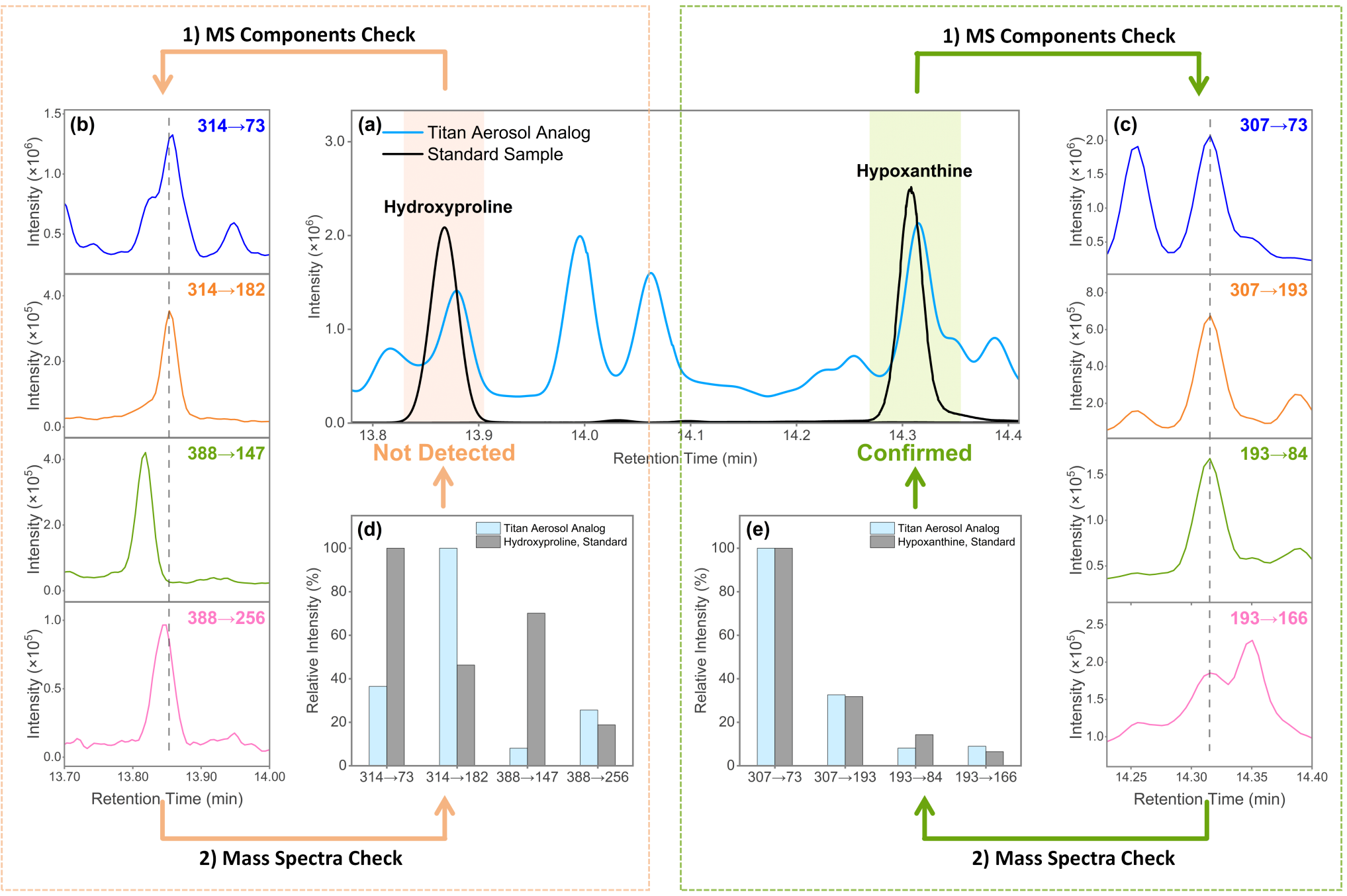

**Figure 2.** Schematic procedure for compound confirmation, illustrated with examples of both confirmed (right half) and excluded (left half) compounds. (a) Total ion chromatograms (TIC) of the Titan aerosol analog (blue line) and standard samples (black line). (b-c) The mass components at corresponding retention times for hydroxyproline and hypoxanthine. (d-e) Comparison of the relative intensities of four ion pairs between the sample and the standards.

Quantification of the compounds was also performed using the same processing method. Taking hypoxanthine as an example, the 307/73 ion pair (Figure 2 (c)) was used as the quantifier. Then we ran a series of standard solutions with concentrations ranging from 1/100000 to 1/50 of the previously prepared solutions, and selected data points with peak areas comparable to those observed in the samples to make calibration curves (peak area versus concentration, Figure 3). By fitting the blank-subtracted peak area of the target compound in the sample to the calibration curve, we estimated its concentration. Using this approach, calibration curves were made for all detected compounds in the two samples (Figure A1–A4), and their concentrations were subsequently calculated (Table A2). The reported concentration uncertainties reflect the combined contributions of GC peak area uncertainties and calibration curve uncertainties.

Compared with GC/MS/MS approaches applied to Bennu samples, our method differs primarily in the analytical methodology. For qualitative identification, we adopted more detailed and robust evaluation criteria to more effectively exclude false-positive results. For quantitative measurements, our work establishes a unified GC/MS/MS framework for the analysis of multiple classes of prebiotic molecules within a single analytical platform. In contrast, GC/MS/MS quantitative measurements reported for Bennu samples have generally focused on individual molecular classes—for example, sugars in Furukawa et al. (2025) and carboxylic acids in Glavin et al. (2025). In addition, the calibration strategy is also different. Furukawa et al. (2025) performed quantitative measurements with GC/MS/MS using an internal standard–based approach. This strategy is well suited for sugars because they share very similar molecular structures and often exhibit comparable derivatization behavior and instrumental response factors. In contrast, the compounds targeted in our study display much larger variability in derivatization efficiency and mass spectrometric response; therefore, our method relies on compound-specific calibration curves to obtain more reliable quantitative results across chemically diverse compound classes.

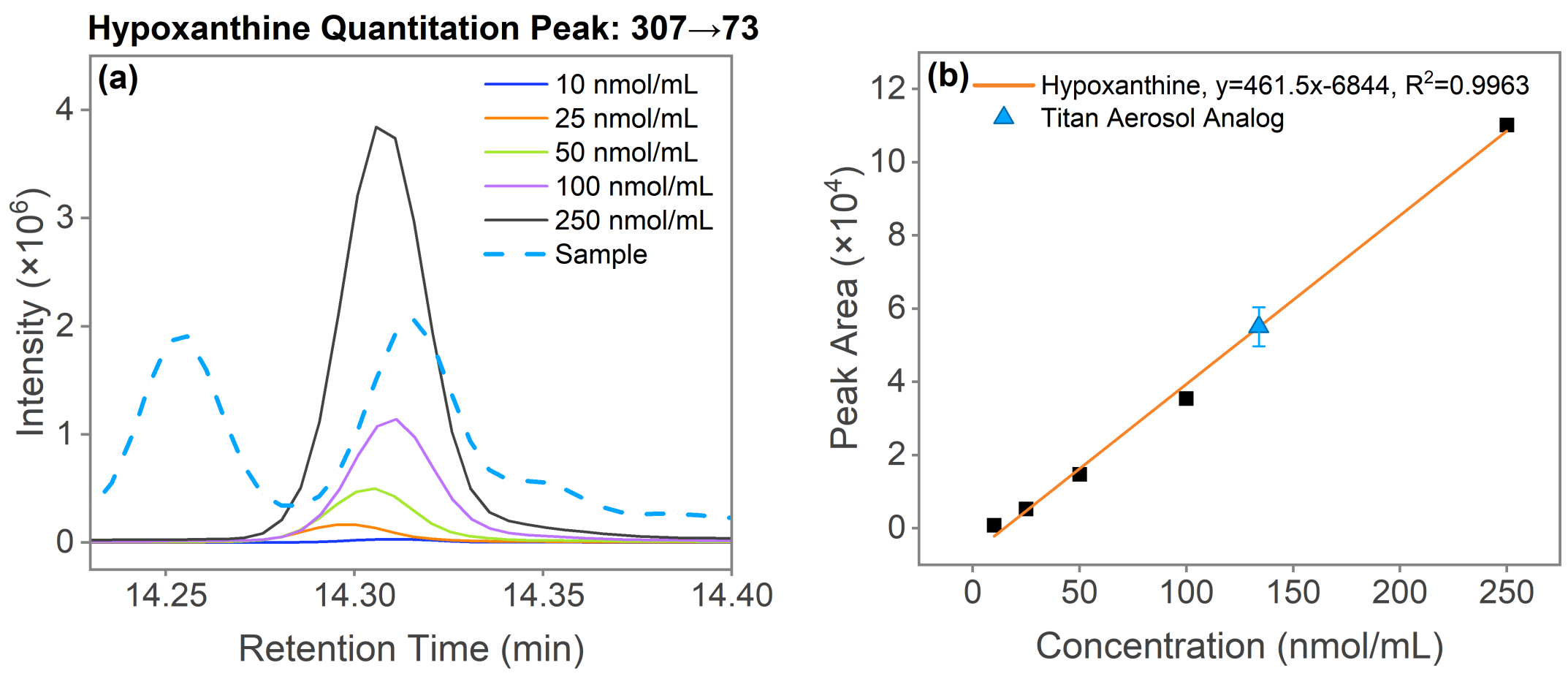

**Figure 3.** Quantitative analysis of hypoxanthine in the Titan aerosol analog. (a) Chromatograms of the selected quantitation peak for the Titan aerosol analog sample and a series of hypoxanthine standard solutions. (b) Corresponding calibration curve for hypoxanthine.

## 3. Results

Applying the established GC/MS/MS quantification method for 56 target molecules (24 amino acids, 7 nucleobases, 21 fatty acids, and 4 other biomolecules), we systematically surveyed these biomolecules in both the Titan aerosol analog and the Martian analog sample. In the Titan aerosol analog, we detected 9 amino acids, 7 nucleobases, 18 fatty acids and 3 other biomolecules (glycolic acid, urea, and guanidine), while in the Martian analog sample, we identified 9 amino acids (6 of them were also found in the Titan aerosol analog), 5 nucleobases, 17 out of 18 fatty acids, and 2 other biomolecules (glycolic acid and urea). Overall, concentrations of the biomolecules are substantially higher in the Titan aerosol analog than in the Martian analog sample. This contrast is expected, as the aerosol analog consists of predominantly organic material, whereas the Martian analog sample is an inorganic mineral matrix containing only trace amounts of organics. Within the Titan aerosol analog, nucleobases generally exhibit the highest concentrations among the three classes of biomolecules, while amino acids are present at the lowest levels. In contrast, in the Martian analog sample, fatty acids are generally 1–2 orders of magnitude more abundant than both amino acids and nucleobases. To further resolve these trends at the compound level, we present below detailed concentration distributions for amino acids (Figure 4), nucleobases (Figure 5), and fatty acids (Figure 6).

### 3.1. Amino Acids

Figure 4 shows the concentrations of individual amino acids detected in the two samples. Among the detected amino acids in the Titan aerosol analog, glycine and proline exhibit the highest and lowest concentrations, respectively, at approximately 12 ppm and 0.1 ppm. The other eight amino acids have concentrations ranging from ~1 to 4 ppm. Glycine, the simplest amino acid, accounts for ~50 wt% of the total amino acid inventory. No amino acids containing more than five carbon atoms are detected, with the exceptions of phenylalanine and tyrosine, both of which contain a benzene ring. Benzene has been identified in Titan's atmosphere (Coustenis et al. 2003) and is considered an important reactant in aerosol formation based on theoretical studies and laboratory simulations (Lebonnois 2005; Trainer et al. 2013), suggesting that this aromatic structure may act as a single reactive unit during amino acid formation. Accordingly, when evaluating amino acid distributions as a function of structural complexity, the benzene ring is treated as a single structural unit; phenylalanine and tyrosine are therefore grouped with C4 amino acids in Figure 4. The concentration pattern of amino acids observed in the Titan aerosol analogs exhibits a strong preference toward structurally simpler molecules, a distribution consistent with formation via abiotic chemical processes. In contrast, amino acid concentrations in the Martian analog sample are relatively uniform, ranging from ~0.3 to 1.3 ppb, with no systematic decrease in abundance with increasing molecular complexity. Notably, three C5-C6 amino acids that are absent from the Titan aerosol analog, valine, leucine and isoleucine, have even higher concentrations than glycine (C2) and serine (C3) in the Martian analog sample. Such a concentration pattern suggests a biotic source of the amino acids in this sample.

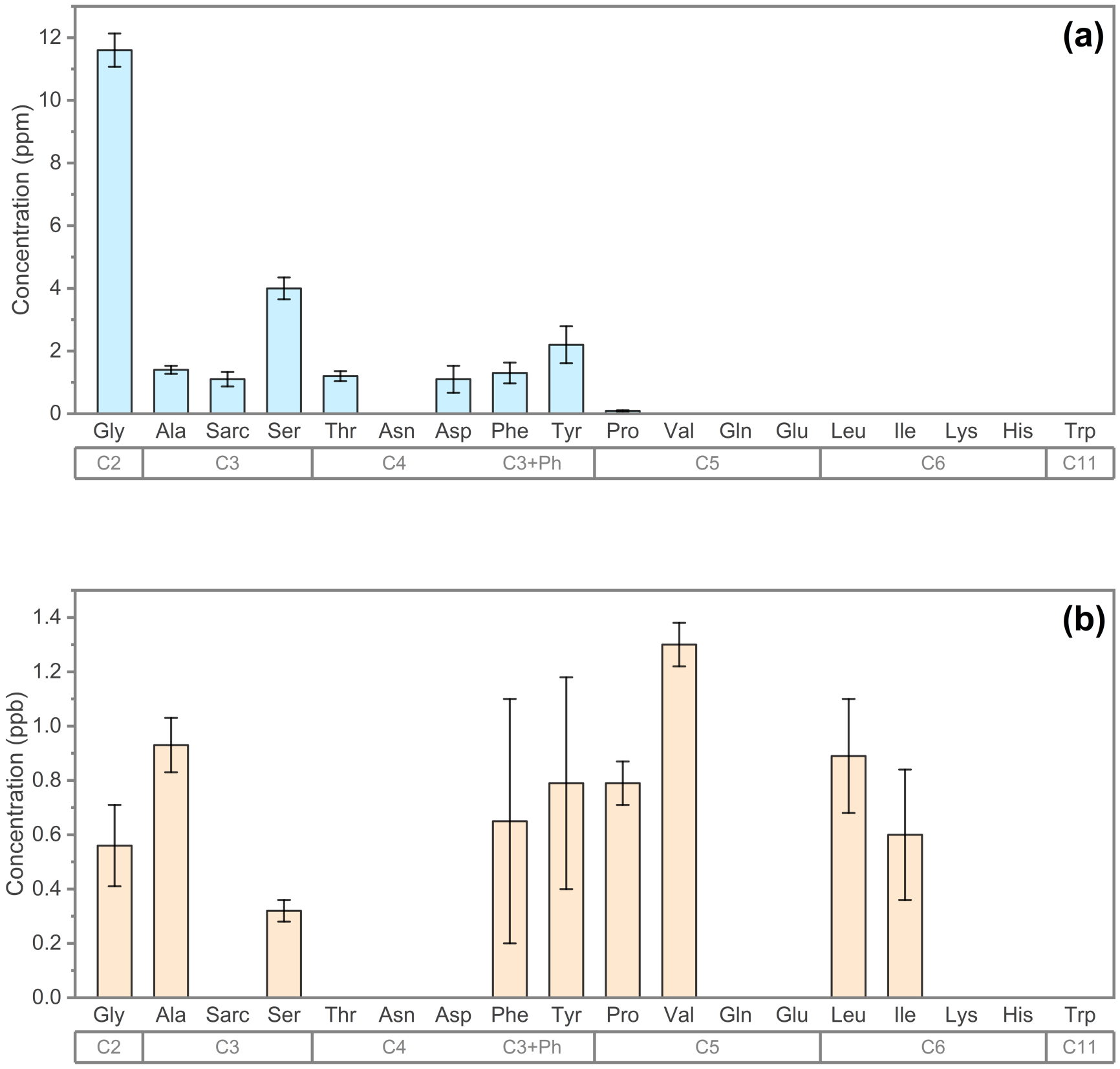

**Figure 4.** Concentrations of amino acids detected in (a) the Titan aerosol analog and (b) the Martian analog sample, arranged in order of increasing structural complexity. Amino acids are grouped by carbon atom number (C2-C11), shown as boxed labels along the x-axis. C3+Ph denotes the two aromatic amino acids, phenylalanine and tyrosine (Ph = phenyl group).

### 3.2. Nucleobases

Figure 5 presents the concentrations of individual nucleobases measured in the two samples. In the Titan aerosol analog, nucleobases are detected at relatively high concentrations ($10^2$–$10^3$ ppm), indicating efficient formation of nitrogen-containing heterocycles under the simulated Titan atmospheric condition. Thymine represents a notable exception, with the lowest concentration at ~7 ppm, potentially reflecting the reduced formation efficiency associated with the presence of a methyl substituent on the heterocyclic ring. A systematic difference is observed between the two major nucleobase classes: pyrimidines (uracil, thymine, cytosine) generally occur at concentrations approximately one order of magnitude lower than purines (hypoxanthine, adenine, xanthine, guanine). This disparity may reflect a limited availability of oxygen-bearing precursors in the experimental gas mixture, which contained only 0.2% CO. By comparison, nucleobase abundances in the Martian analog sample are markedly lower and display greater heterogeneity. Guanine is the most abundant nucleobase (~340 ppb), followed by hypoxanthine (~20 ppb). The other three detected nucleobases have concentrations comparable to those of amino acids (~0.3 to 2 ppb) in this sample, while cytosine and xanthine are not detected. Together, these results highlight a pronounced divergence in both the absolute abundances and relative distributions of nucleobases between the two samples, reflecting fundamental differences in organic inventory and the formation or preservation processes.

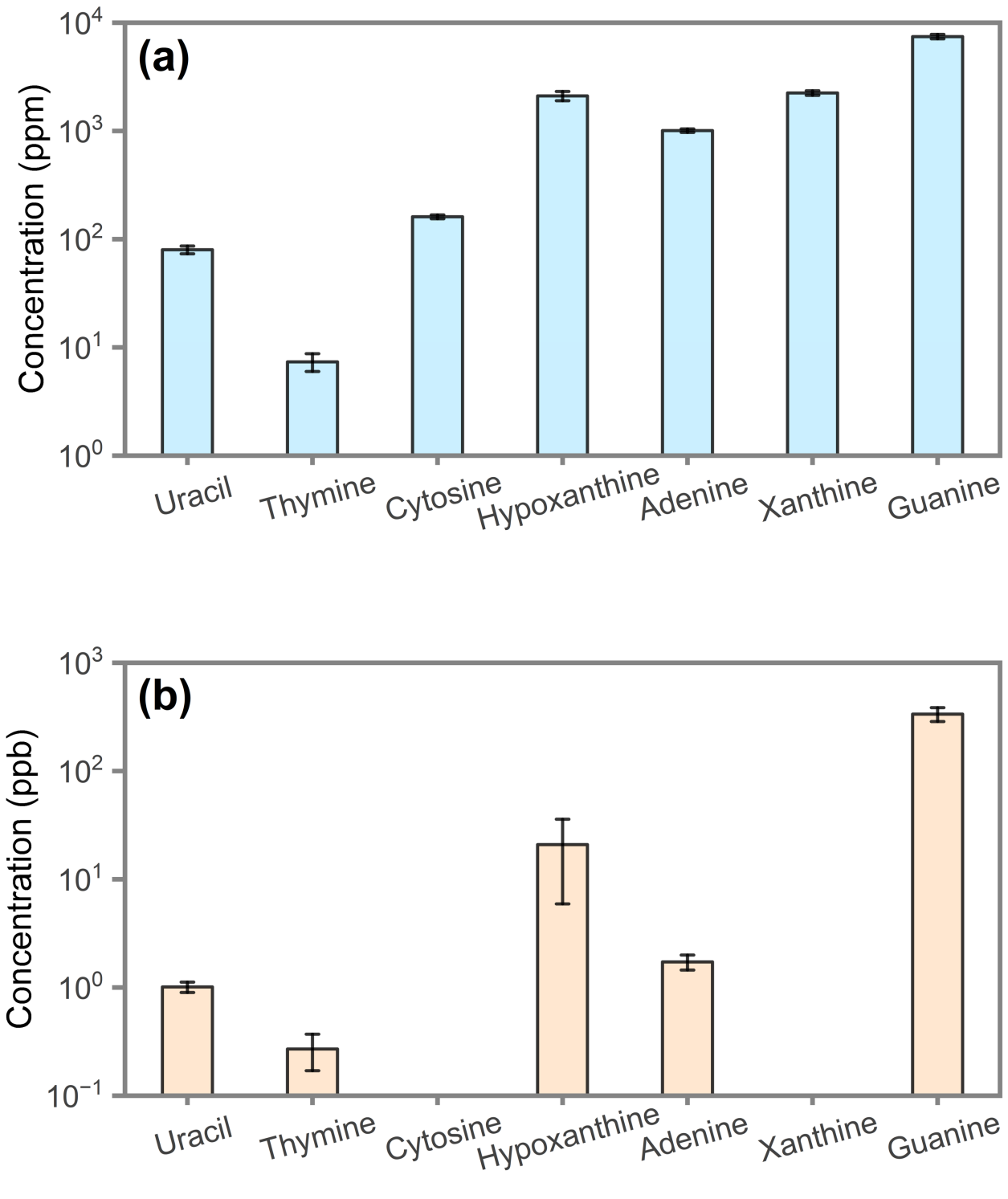

**Figure 5.** Concentrations of nucleobases detected in (a) the Titan aerosol analog and (b) the Martian analog sample, arranged in order of chromatographic retention time.

### 3.3. Fatty Acids

Figure 6 summarizes the concentration distributions of the fatty acids in the two samples, including 15 saturated straight-chain fatty acids (from C12:0 to C26:0) and 3 unsaturated fatty acids (C16:1 n-7, C18:2 n-6, C18:1 n-9). In the Titan aerosol analog sample, fatty acid abundances span nearly three orders of magnitude. The most prominent components are C16:0 (~300 ppm) and C18:0 (~240 ppm), followed by C16:1 n-7 (~130 ppm), and the long-chain species C26:0 (~110 ppm). The remaining fatty acids have lower concentrations ranging from ~1 to 40 ppm, with a predominance of even-numbered carbon chains. Fatty acid inventories in the Martian analog sample exhibit substantially lower absolute abundances. The dominant species are C16:0 (3.6 ppm), C18:1 n-9 (2.7 ppm) and C18:0 (0.8 ppm), whereas C12:0, C13:0, C19:0, and C21:0 are present at the lowest levels (~20–60 ppb), and C25:0 is not detected. The remaining fatty acids fall within an intermediate range of ~0.1 to 0.5 ppm. Despite the much lower abundance, the preference for even-carbon-numbered chains persists in this sample. The fatty acid results reveal systematic differences in both absolute concentrations and relative distributions between the two analog materials, while sharing similar even-over-odd carbon number dominance—that alone does not distinguish between abiotic and biological origins.

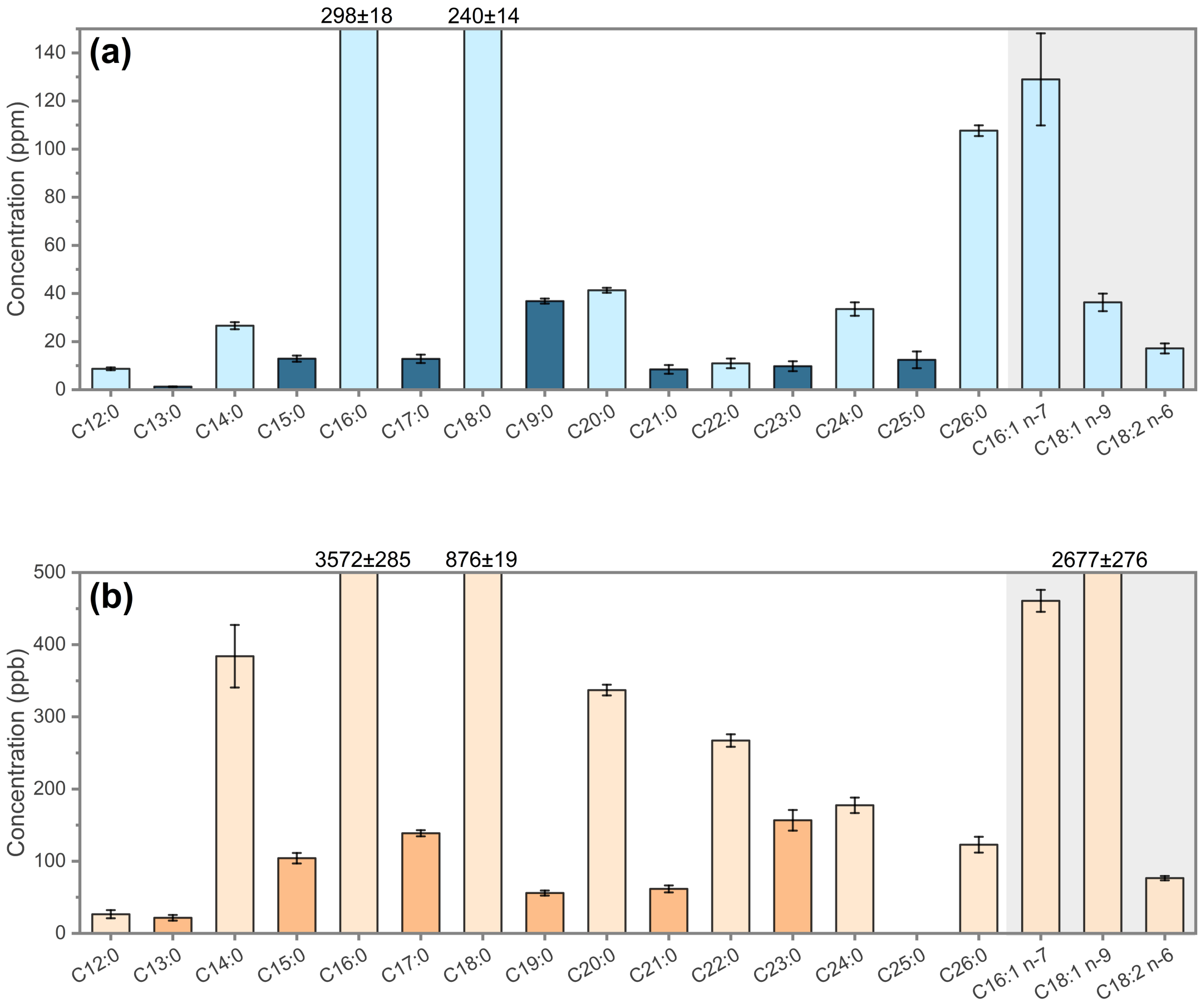

**Figure 6.** Concentrations of fatty acids detected in (a) the Titan aerosol analog and (b) the Martian analog sample. Fatty acids with odd and even numbers of carbon atoms are represented by darker and lighter bars, respectively, and arranged in order of increasing carbon atom number. All are straight-chain saturated fatty acids, except three unsaturated fatty acids (grouped in the gray region on the right).

## 4. Discussion

The GC/MS/MS framework developed in this study expands the target compound list to include fatty acids and implements an improved data-processing workflow to enable more robust compound identification and quantification. These methodological advances allow reliable comparisons of molecular abundance patterns across different environments, which are critical for interpreting prebiotic chemistry and assessing potential biosignatures. Applying this approach to two representative planetary analog materials, we obtained a comprehensive view of the molecular inventories and concentration patterns of biomolecules. Next, we explore the broader implications of our findings from the perspectives of prebiotic chemistry and biosignature identification.

### 4.1. Implications for Prebiotic Chemistry

In the Titan aerosol analog sample, we observed the formation of a wide variety of organic building blocks essential for prebiotic chemistry. Because we used a CO mixing ratio much higher than the actual value, our results do not represent biomolecule production in Titan's atmosphere but reveal the potential of biomolecule formation under related atmospheric conditions. Notably, we report for the first time the presence and concentration distributions of long-chain fatty acids (C12–C26) in an organic aerosol analog. The relatively high abundances of these fatty acids were unexpected, as most nonbiological samples, such as products from Miller-Urey experiments (Miller 1955; Allen & Ponnamperuma 1967; Yuen et al. 1981) and carbonaceous meteorites (Lai et al. 2019), typically do not contain fatty acids with chain length exceeding C12. Although several studies reported the presence of longer-chain fatty acids in some meteorites (Nagy et al. 1963; Smith & Kaplan 1970), these occurrences are commonly attributed to terrestrial contamination. A similar possibility must also be considered in the present study, as long-chain fatty acids are ubiquitous in laboratory environments and contamination therefore cannot be completely excluded. While procedural blanks were analyzed, they were not carried through all solid-handling steps that may introduce such compounds, including sample collection, weighing, and storage. If these fatty acids are indeed formed during aerosol production rather than introduced by contamination, their efficient production of C12–C26 fatty acids through atmospheric photochemistry would have important implications for prebiotic chemistry. Fatty acids with chain lengths of at least eight carbon atoms are required to form stable vesicles (Apel et al. 2002), and the availability of longer-chain amphiphiles could therefore facilitate the spontaneous formation and persistence of protocell-like membrane structures upon deposition of organic aerosols onto planetary surfaces.

In addition to fatty acids, we observed all seven targeted nucleobases in the Titan aerosol analog. The nucleobases were detected previously in several types of prebiotic materials such as similar organic aerosol analogs (Sebree et al. 2018; Pearce et al. 2024a), Miller-Urey experiment products (Ferus et al. 2017), carbonaceous meteorites (Callahan et al. 2011), and returned asteroid samples (Mojarro et al. 2025). However, the nucleobase concentrations observed here are approximately one to two orders of magnitude higher than those reported in earlier studies of both the Hadean Earth (Pearce et al. 2024a) and Titan aerosol analogs (Pearce et al. 2024b). This increase is likely attributed to the higher energy density of the ECR plasma used in our experiment (~1800 W), which is roughly ten times greater than that of previous experiments (~170 W, Pearce et al. 2024a), potentially promoting more efficient formation of nitrogen-containing heterocycles. By comparison, nucleobases in meteorites are typically present at much lower levels, ranging from below 1 ppb to a few hundred ppb. The elevated concentrations of nucleobases observed in our study may substantially increase the probability of further chemical evolutions, once such organic aerosols are delivered to surface aqueous environments. Such environments could include transient impact-generated pools on Titan or "warm little ponds" on the Hadean Earth. Based on estimates by Pearce et al. (2024b), nucleobases produced in our simulated conditions could reach concentrations exceeding ~100 μM in such ponds, a level considered sufficient to support nucleotide formation (Ponnamperuma et al. 1963). These results suggest that energetic atmospheric processing may supply nucleobases at concentrations relevant to subsequent prebiotic processes, thereby promoting chemical complexity towards life's origin.

With respect to amino acids, we identified 8 proteinogenic amino acids in the Titan aerosol analog, including phenylalanine, which is detected in organic aerosol analogs for the first time. The concentrations of detected amino acids in our study are comparable with those reported in some earlier studies (Pearce et al. 2024a; Pearce et al. 2024b). At the same time, several other proteinogenic amino acids (valine, leucine, isoleucine, and tryptophan) previously detected in organic aerosol analogs were not observed in our sample. The difference highlights the sensitivity of amino acid inventories to experimental conditions and gas-phase composition. The diversity of amino acids observed in the Titan aerosol analog shows substantial overlap with that observed in carbonaceous meteorites and asteroid returned samples. In particular, CM2 and CR2 meteorites, which exhibit the highest amino acid abundances and diversity among meteoritic samples, contain all 9 proteinogenic amino acids identified here, along with glutamic acid, leucine and isoleucine (Cobb & Pudritz 2014). In addition, the amino acid concentrations in these meteorites are generally within the same range as those in our sample, suggesting atmospheric synthesis and parent-body processing may generate broadly similar amino acid inventories. By comparison, Miller-Urey experiments have yielded a greater diversity of amino acids (Johnson et al. 2008; Parker et al. 2011), including two species containing amide functional groups (glutamine and asparagine), and one sulfur-bearing amino acid (methionine) when $H_2S$ is included in the experiments. The gas mixtures used in these experiments (e.g. $CH_4$, $NH_3$, $H_2S$ and $CO_2$) provide a more strongly reducing environment and a higher availability of oxygen- and sulfur-bearing precursors, which are likely more favorable for amino acid synthesis. Furthermore, the explicit inclusion of liquid water in Miller-Urey-type experiment enables Strecker-type reactions, which require water as a reactant and are known to efficiently produce α-amino acids (Miller 1957).

The combined results for fatty acids, nucleobases, and amino acids demonstrate that energetic atmospheric processing can generate a diverse inventory of prebiotically relevant species, while also imposing distinct molecular and structural biases on their relative abundances. The coexistence of membrane-forming amphiphiles, nucleobases, and amino acid monomers at measurable concentrations underscores the potential of atmospheric chemistry to seed subsequent stages of prebiotic chemical evolution once these materials are delivered to aqueous environments. At the same time, the differences observed among molecular classes and between planetary analogs highlight the importance of environmental context—including redox state, precursor availability, and post-depositional processing—in shaping prebiotic chemical pathways.

For the Hadean Earth, atmospheric composition remains poorly constrained. Current consensus suggests that degassing from Earth's mantle would have produced a $CO_2$–$N_2$–dominated atmosphere (Shaw 2008), intermittently modified by impact-driven, more reducing episodes enriched in $H_2$ and CO, with minor contributions of $CH_4$ and $NH_3$ (Zahnle et al. 2020). Sulfur-bearing gases such as $SO_2$ and $H_2S$ could also have been supplied through degassing processes (Marchi et al. 2016). Consequently, investigations of prebiotic processes on the Hadean Earth should consider more complex gas mixtures to assess the atmospheric production of structurally diverse biomolecules. Beyond atmospheric synthesis alone, the subsequent evolution of organic aerosols in small bodies of water, such as ponds or shallow pools, warrants further study to assess how biomolecule diversity and quantity change during aqueous processing, and to determine whether the periodic dry-wet cycles can concentrate biomolecules and facilitate polymerization.

For Titan, the atmospheric synthesis of biomolecules is likely limited by the low availability of oxygen-bearing species and may occur at abundances far lower than those detected in our aerosol analog sample. Nevertheless, aqueous alterations may still take place in transient liquid water environments on the surface, such as impact-generated melt pools or cryovolcanic flows (O'Brien et al. 2005; Neish et al. 2006), potentially increasing both the diversity and abundance of biomolecules. To better understand these processes on Titan, future studies should simulate conditions that more closely resemble Titan's atmospheric environment, with tighter control over experimental parameters, such as gas compositions, pressure, temperature, and energy input. The use of isotopically labeled precursor gases (e.g., $^{13}C$-labeled gases) during aerosol analog production would also be needed to distinguish between compounds formed in the simulation experiments and those arising from laboratory contamination.

### 4.2. Implications for Biosignatures

The gypsum we analyzed as a Martian analog sample, as reported in a previous study (Li et al. 2024), was originally deposited in fractures caused by the uplift of the salt domes, during the upward migration of deep fluids. Methane-dominated hydrocarbons from deep in the Qaidam Basin have been detected in these gypsum crystals. This, combined with our detection of biomolecules, may indicate the preservation of signatures from a deep subsurface biosphere. Given that features resembling gypsum ridges in the Qaidam Basin have been observed on Mars (Rapin et al. 2016), it is plausible that biosignatures from potential habitable zones in the Martian subsurface could also be preserved in such crystals. Therefore, applying the GC/MS/MS method to Martian analog samples is valuable for future Mars sample return missions, offering an approach for searching for biosignatures on Mars.

Our results of the Martian analog sample allow an assessment of which detected biomolecules may constitute more robust biosignature candidates. A key criterion for biosignatures is their detectability, defined by the requirement that sources exceed sinks over geologic timescales. In the case of our Martian analog sample, sinks refer to degradation pathways such as radiolysis and chemical alteration, whereas sources depend on the biological or geochemical production of these biomolecules. Biomolecules that are chemically more stable and common across diverse life forms are more likely to persist and remain detectable for longer periods, thus increasing their chances of being noticed. Consistent with this expectation, we observe a clear relationship between compound stability and measured abundance. Fatty acids, being generally more stable than amino acids and nucleobases, are found at much higher concentrations. Among the nine detected amino acids, seven have the most stable nonpolar R groups (glycine, alanine, valine, leucine, isoleucine, proline, phenylalanine), two have more reactive polar uncharged R groups (serine, tyrosine), whereas none of the amino acids with charged R groups are detected. Similarly, among the seven targeted nucleobases, the absence of cytosine and xanthine may reflect their lower stability, with cytosine prone to deamination to uracil (Lindahl 1996) and xanthine destabilized by the presence of two carbonyl groups. Although the original sources of these biomolecules cannot be uniquely constrained, several notable abundance patterns are consistent with biological influences. For example, guanine exhibits a particularly high concentration relative to other nucleobases, which may be explained by its widespread occurrence as biogenic guanine crystals in various life forms, often serving as a waste or storage product of nitrogen metabolism in microbes (Mojzeš et al. 2020). Similarly, hypoxanthine, another abundant nucleobase, is also an important metabolite and can be accommodated within guanine crystals (Pinsk et al. 2022). For fatty acids, the dominance of C16:0 and C18:0 aligns with their prevalence in cellular membranes across a wide range of organisms. In addition, the observed even-over-odd carbon number predominance in fatty acid distributions is consistent with microbial cellular metabolic processes, where biosynthesis commonly involves the successive incorporation of acetyl (C2) units, leading to a preference for even-numbered carbon chains (Volkman 2005).

If only considering the detectability, fatty acids would appear to be the most promising candidates among the three groups of biomolecules for biosignature detection in Martian samples, owing to their relatively high abundances and greater chemical stability. This assessment is consistent with previous observations of lipid biosignatures in other Martian analog environments on Earth (Tan et al. 2018). However, another important criterion for biosignature identification is the ability to distinguish biological signals from abiotic background sources. In this regard, fatty acids present a more ambiguous case. Abiotic and biotic fatty acids are often believed to exhibit different concentration patterns: biotic fatty acids tend to show relatively irregular distributions with an even-over-odd predominance, whereas abiotic fatty acids typically display monotonically decreasing or unimodal distributions with no systematic carbon-number preference, as observed in meteorites (Lai et al. 2019) and in hydrothermal experiments involving the Fischer-Tropsch-type (FTT) synthesis (Nooner & Oro 1979). However, our Titan aerosol analog sample suggests that abiotic pathways may also generate fatty acid distributions resembling those typically attributed to biological sources, including even-over-odd carbon number dominance. If these fatty acids are not derived from laboratory contamination, such patterns may arise from chain-growth mechanisms through the incorporation of C2 units (e.g., $C_2H_4$ or $C_2H_2$). Additionally, the unexpectedly high yields of C16:0, C18:0, C24:0, and C26:0 suggest that some unconstrained mechanisms may operate in complex atmospheric systems, where gas-phase reactions among radicals, ions, and neutral species coexist with heterogeneous reactions on aerosol particle surfaces. In this context, fatty acid abundance patterns alone may not always provide definitive evidence of biological activity, highlighting the necessity of integrating complementary biosignatures, such as isotopic compositions at specific positions within fatty acids (Monson & Hayes 1982; McCollom & Seewald 2006). Alternatively, if the detected fatty acids originate from laboratory contamination, this likely reflects, at least in part, the difficulty of fully controlling contamination for such compounds. Similar issues may also arise for planetary materials, where fatty acids could be introduced during sample collection and handling. Consequently, even when fatty acids are detected and identified as biologically derived, distinguishing extraterrestrial sources from terrestrial contamination remains challenging.

In contrast, the amino acid concentration patterns in the two samples (Figure 4) are distinct and diagnostically informative, reflecting differences between abiotic and biotic sources. In the Titan aerosol analog, the dominance of simpler amino acids, especially glycine, is consistent with findings from Miller-Urey experiments and meteorite studies. Two detected amino acids with benzene ring (phenylalanine and tyrosine) in the Titan aerosol analog were also observed in meteorites (Pizzarello & Holmes 2009) and returned asteroid samples (Mojarro et al. 2025), confirming that these two aromatic amino acids can form abiotically due to their low structural complexity. Importantly, 20 proteinogenic amino acids span a wide range of functional groups and side-chain chemistries (e.g., -$NH_2$, -COOH, -OH, aromatic rings), giving rise to diverse formation pathways and reaction barriers. Such structural diversity enhances the likelihood that abiotic processes preferentially produce simpler species, whereas biological systems can generate more complex and compositionally balanced amino acid distributions. By comparison, fatty acids consist of linear hydrocarbon chains terminated by a carboxyl group and are less complex structurally, potentially leading to greater overlap between abiotic and biotic production pathways. Therefore, when using concentration pattern as a biosignature criterion, it is more effective to choose a group of biomolecules with greater structural diversity, such as amino acids. On this basis, amino acids are particularly important in the search for biosignatures in Martian samples, as they can provide more compelling evidence for life. Our results of the Martian analog sample indicate that although the amino acids are relatively less stable than fatty acids, they can nonetheless be preserved over geologically significant timescales under favorable conditions against degradation. This interpretation is also supported by a recent study suggesting that amino acids trapped in Martian surface ice could survive over 50 million years of cosmic ray radiation (Pavlov et al. 2025). Building on these insights, future work will extend our analytical method to a broader range of mineral and sediment samples to assess their potential for preserving biosignatures. In parallel, given the limited quantity and low biomolecule concentrations anticipated for future returned Martian samples, further refinement of sample preparation protocols will be essential, with particular emphasis on contamination control and efficient extraction of trace biomolecules.

## 5. Conclusion

This study confirms that gas chromatography–tandem mass spectrometry (GC/MS/MS) provides the sensitivity and selectivity required for robust identification and quantification of trace building blocks of life in complex planetary matrices. Application of an optimized GC/MS/MS approach to a Titan aerosol analog and a Martian gypsum analog from the Qaidam Basin reveals diverse inventories of amino acids, nucleobases, and fatty acids at ppb to ppm levels, illustrating its broad analytical capability.

In the Titan aerosol analog, the simultaneous presence of ten amino acids, an unexpectedly broad range of long-chain fatty acids, and elevated nucleobase abundances reveals that energetic atmospheric photochemistry can generate a chemically diverse inventory of prebiotically relevant molecules. These results suggest that organic aerosols may act as an efficient source for delivering complex mixtures of these molecules to surface or near-surface aqueous environments, where further chemical evolution may occur. In the Martian analog sample, the presence of trace amino acids and nucleobases indicates that even relatively labile biomolecules may persist over geologically significant timescales when protected by mineral matrices under Mars-relevant conditions. More importantly, comparison between the two samples highlights fundamental differences in the diagnostic value of molecular classes for biosignature identification. Fatty acid features often considered indicative of biological activity—such as even-carbon-number predominance and the relative abundance of C16 and C18 homologs—are also observed in the Titan aerosol analog. If these detections are genuine, it suggests that such characteristics alone may not provide unambiguous evidence for a biological origin. Conversely, if they arise from laboratory contamination, the results highlight the challenges of fully controlling biological contamination and underscore the need for extreme caution when analyzing these materials. In contrast, amino acid abundance patterns show clear and consistent distinctions between abiotic and biotic sources across the two samples, reinforcing their robustness as biosignature candidates. Together, these findings refine current perspectives on molecular biosignatures and emphasize the importance of considering both molecular stability and structural diversity when evaluating potential biological signals. This work underscores the value of quantitative, pattern-based analyses for advancing our understanding of prebiotic chemical evolution and provides a framework for future biosignature investigations in planetary exploration, particularly in the context of Mars sample return missions.

**Acknowledgments**

The authors gratefully acknowledge the support from the National Key Research and Development Program of China (2024YFF0809800) and the National Natural Science Foundation of China (42475132). We thank Dr. Ben K. D. Pearce and the anonymous reviewer whose comments helped to significantly improve the manuscript.

## Appendix

**Table A1.** Retention times and selected reaction monitoring (SRM) transitions used for GC/MS/MS detection of biomolecules in this study.

| Biomolecule | Retention Time (min) | Parent Ions (m/z) | Daughter Ions (m/z) | Collision Energies (eV) |
|---|---|---|---|---|
| Amino Acids | | | | |
| Alanine ($C_3H_7NO_2$) | 7.34 | 158<br>232 | 73, 102<br>73, 147 | 6, 12<br>32, 10 |
| Glycine ($C_2H_5NO_2$) | 7.64 | 202<br>218 | 132, 160<br>131, 147 | 10, 6<br>24, 10 |
| Sarcosine ($C_3H_7NO_2$) | 8.03 | 232<br>260 | 131, 147<br>158, 232 | 26, 10<br>10, 6 |
| Valine ($C_5H_{11}NO_2$) | 8.64 | 186<br>260 | 73, 130<br>73, 147 | 14, 6<br>36, 12 |
| Leucine ($C_6H_{13}NO_2$) | 9.08 | 200<br>274 | 73, 144<br>73, 147 | 14, 6<br>36, 12 |
| Isoleucine ($C_6H_{13}NO_2$) | 9.43 | 200<br>274 | 73, 144<br>73, 147 | 14, 6<br>36, 12 |
| Proline ($C_5H_9NO_2$) | 9.86 | 184<br>258 | 73, 128<br>73, 147 | 14, 6<br>34, 12 |
| Methionine ($C_5H_{11}NO_2S$) | 11.86 | 218<br>292 | 73, 170<br>147, 244 | 16, 6<br>14, 6 |
| Serine ($C_3H_7NO_3$) | 12.03 | 288<br>362 | 73, 100<br>147, 230 | 16, 12<br>18, 6 |
| Threonine ($C_4H_9NO_3$) | 12.31 | 303<br>404 | 73, 202<br>244, 376 | 20, 12<br>8, 8 |
| Phenylalanine ($C_9H_{11}NO_2$) | 13.02 | 302<br>336 | 73, 218<br>147, 308 | 18, 6<br>20, 8 |
| Aspartic Acid ($C_4H_7NO_4$) | 13.57 | 390<br>418 | 147, 216<br>147, 390 | 16, 8<br>22, 6 |
| Hydroxyproline ($C_5H_9NO_3$) | 13.86 | 314<br>388 | 73, 182<br>147, 256 | 22, 8<br>18, 8 |
| Cysteine ($C_3H_7NO_2S$) | 13.99 | 378<br>406 | 147, 246<br>147, 378 | 18, 6<br>26, 8 |
| Glutamic Acid ($C_5H_9NO_4$) | 14.62 | 330<br>432 | 73, 170<br>147, 272 | 20, 8<br>22, 8 |
| Ornithine ($C_5H_{12}N_2O_2$) | 14.66 | 184<br>286 | 73, 128<br>147, 258 | 12, 6<br>12, 8 |
| Asparagine ($C_4H_8N_2O_3$) | 14.87 | 244<br>417 | 73, 133<br>147, 232 | 18, 12<br>22, 8 |
| Lysine ($C_6H_{14}N_2O_2$) | 15.50 | 198<br>300 | 73, 142<br>168, 272 | 14, 8<br>8, 8 |
| Aminoadipic Acid ($C_6H_{11}NO_4$) | 15.52 | 344<br>446 | 73, 301<br>147, 286 | 24, 8<br>22, 8 |
| Glutamine ($C_5H_{10}N_2O_3$) | 15.84 | 385<br>431 | 147, 253<br>357, 385 | 18, 8<br>14, 10 |
| Histidine ($C_6H_9N_3O_2$) | 17.24 | 224<br>440 | 73, 110<br>280, 412 | 18, 12<br>10, 10 |
| Tyrosine ($C_9H_{11}NO_3$) | 17.57 | 302<br>466 | 73, 218<br>147, 438 | 18, 6<br>28, 10 |
| Tryptophan ($C_{11}H_{12}N_2O_2$) | 20.28 | 244<br>302 | 73, 188<br>73, 218 | 16, 8<br>16, 6 |
| Cystine ($C_6H_{12}N_2O_4S_2$) | 25.10 | 258<br>348 | 147, 230<br>73, 302 | 16, 6<br>20, 8 |
| | | Nucleobases | | |
| Uracil ($C_4H_4N_2O_2$) | 9.90 | 99<br>283 | 69, 71<br>99, 241 | 28, 12<br>16, 8 |
| Thymine ($C_5H_6N_2O_2$) | 10.58 | 113<br>297 | 59, 85<br>73, 113 | 12, 6<br>24, 16 |
| Cytosine ($C_4H_5N_3O$) | 11.80 | 212<br>282 | 73, 170<br>98, 212 | 16, 10<br>22, 12 |
| Hypoxanthine ($C_5H_4N_4O$) | 14.31 | 193<br>307 | 84, 166<br>73, 193 | 24, 8<br>18, 14 |
| Adenine ($C_5H_5N_5$) | 15.04 | 192<br>306 | 84, 165<br>73, 192 | 24, 10<br>18, 12 |
| Xanthine ($C_5H_4N_4O_2$) | 17.28 | 265<br>437 | 131, 158<br>147, 363 | 22, 12<br>18, 12 |
| Guanine ($C_5H_5N_5O$) | 18.15 | 264<br>436 | 131, 158<br>147, 322 | 22, 14<br>20, 14 |
| | | Fatty Acids | | |
| Dodecanoic Acid ($C_{12}H_{24}O_2$) | 11.15 | 129<br>257 | 75, 85<br>75, 131 | 10, 6<br>16, 8 |
| Tridecanoic Acid ($C_{13}H_{26}O_2$) | 12.13 | 129<br>271 | 75, 85<br>75, 131 | 10, 6<br>16, 8 |
| Isostearic Acid ($C_{18}H_{36}O_2$) | 12.36 | 129<br>341 | 75, 85<br>75, 229 | 10, 6<br>22, 8 |
| Myristic Acid ($C_{14}H_{28}O_2$) | 13.08 | 129<br>285 | 75, 85<br>75, 131 | 10, 6<br>18, 8 |
| Pentadecanoic Acid ($C_{15}H_{30}O_2$) | 13.99 | 129<br>299 | 75, 85<br>75, 131 | 10, 6<br>16, 8 |
| Palmitoleic Acid ($C_{16}H_{30}O_2$) | 14.70 | 129<br>311 | 75, 85<br>75, 121 | 10, 6<br>18, 8 |
| Palmitic Acid ($C_{16}H_{32}O_2$) | 14.86 | 129<br>313 | 75, 85<br>75, 131 | 10, 6<br>18, 8 |
| Heptadecanoic Acid ($C_{17}H_{34}O_2$) | 15.70 | 129<br>327 | 75, 85<br>75, 131 | 10, 6<br>16, 10 |
| γ-Linolenic Acid ($C_{18}H_{30}O_2$) | 16.09 | 129<br>335 | 75, 85<br>75, 159, 243 | 10, 6<br>24, 10, 6 |
| Linoleic Acid ($C_{18}H_{32}O_2$) | 16.27 | 129<br>337 | 75, 85<br>75, 105 | 10, 6<br>20, 10 |
| Oleic Acid ($C_{18}H_{34}O_2$) | 16.31 | 129<br>339 | 75, 85<br>75, 121 | 10, 6<br>20, 8 |
| α-Linolenic Acid ($C_{18}H_{30}O_2$) | 16.38 | 129<br>335 | 75, 85<br>75, 131, 145 | 12, 6<br>24, 10, 8 |
| Stearic Acid ($C_{18}H_{36}O_2$) | 16.51 | 129<br>341 | 75, 85<br>75, 131 | 10, 6<br>18, 10 |
| Nonadecanoic Acid ($C_{19}H_{38}O_2$) | 17.33 | 129<br>355 | 75, 85<br>75, 131 | 12, 6<br>20, 10 |
| Arachidic Acid ($C_{20}H_{40}O_2$) | 18.18 | 129<br>369 | 75, 85<br>75, 131 | 12, 8<br>20, 12 |
| Heneicosanoic Acid ($C_{21}H_{42}O_2$) | 19.20 | 129<br>383 | 75, 85<br>95, 131 | 12, 6<br>12, 10 |
| Behenic Acid ($C_{22}H_{44}O_2$) | 20.45 | 129<br>397 | 75, 85<br>95, 131 | 12, 6<br>12, 12 |
| Tricosanoic Acid ($C_{23}H_{46}O_2$) | 21.96 | 129<br>411 | 75, 85<br>95, 131 | 12, 6<br>12, 12 |
| Tetracosanoic Acid ($C_{24}H_{48}O_2$) | 23.86 | 129<br>425 | 75, 85<br>95, 131 | 12, 8<br>14, 14 |
| Pentacosanoic Acid ($C_{25}H_{50}O_2$) | 26.23 | 129<br>439 | 75, 85<br>95, 131 | 12, 8<br>14, 14 |
| Hexacosanoic Acid ($C_{26}H_{52}O_2$) | 29.16 | 129<br>453 | 75, 85<br>95, 131 | 12, 8<br>16, 14 |
| Other Biomolecules | | | | |
| Glycolic Acid ($C_2H_4O_3$) | 7.00 | 189<br>247 | 73, 147<br>147, 189 | 28, 8<br>12, 6 |
| Urea ($CH_4N_2O$) | 8.70 | 231<br>274 | 147, 173<br>147, 218 | 10, 8<br>18, 6 |
| Guanidine ($CH_5N_3$) | 11.77 | 213<br>344 | 73, 171<br>146, 255 | 22, 8<br>18, 6 |
| Cystathionine ($C_7H_{14}N_2O_4S$) | 23.22 | 302<br>362 | 73, 218<br>73, 202 | 18, 6<br>22, 8 |

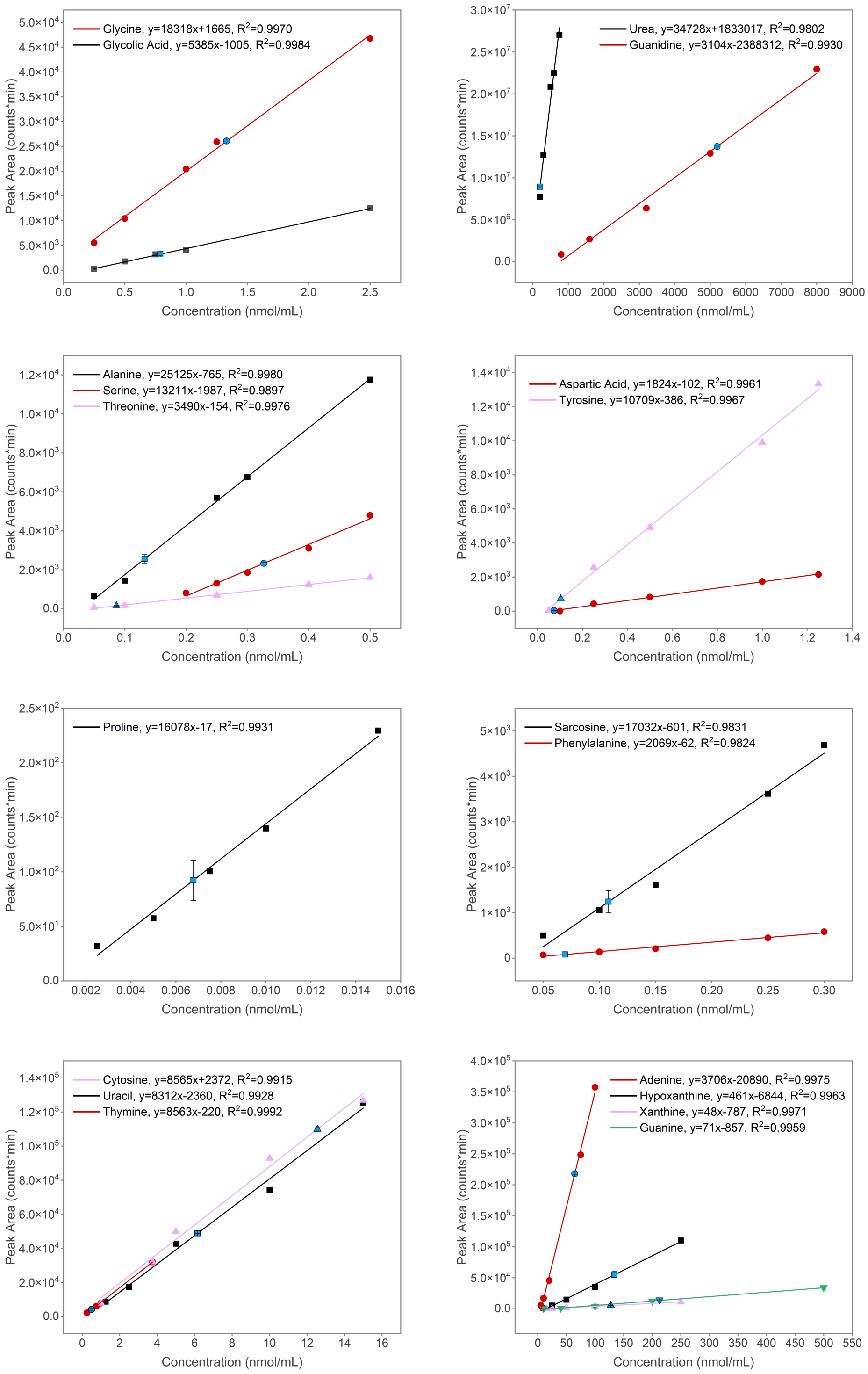

**Figure A1.** Calibration curves for the 10 amino acids, 7 nucleobases, and 3 other biomolecules detected in the Titan aerosol analog, with blue symbols representing data points obtained from the Titan aerosol analog sample. Calibration points were selected to match the corresponding signal intensities, resulting in substantially different concentration

ranges among compounds in Figures A1–A4. Note that the concentration values were converted to represent the concentrations of biomolecules in the original sample solutions (Figure A1 and A2 correspond to the Titan aerosol analog, while Figure A3 and A4 correspond to the Martian analog sample).

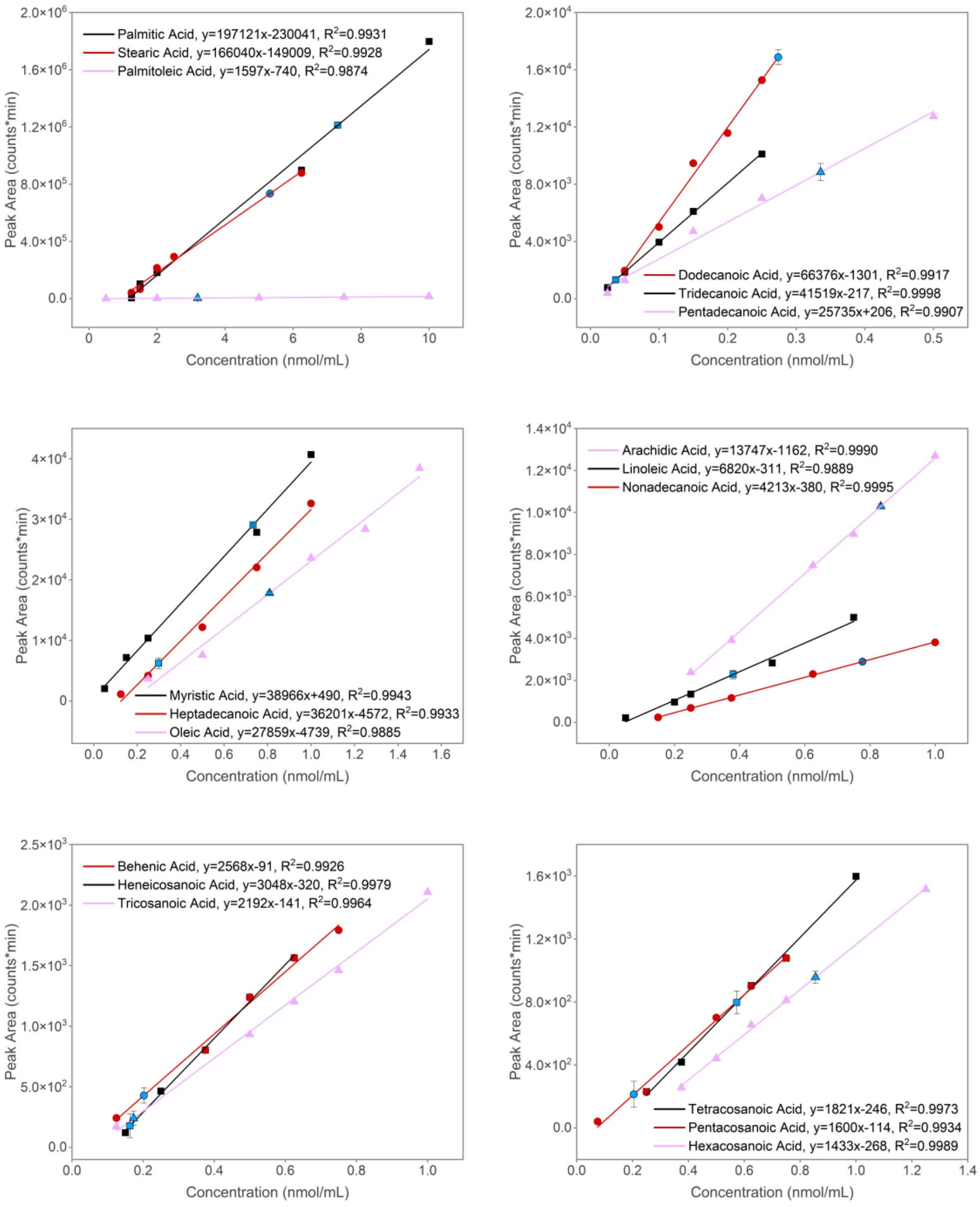

**Figure A2.** Calibration curves for the 18 fatty acids detected in the Titan aerosol analog, with blue symbols representing data points obtained from the Titan aerosol analog sample.

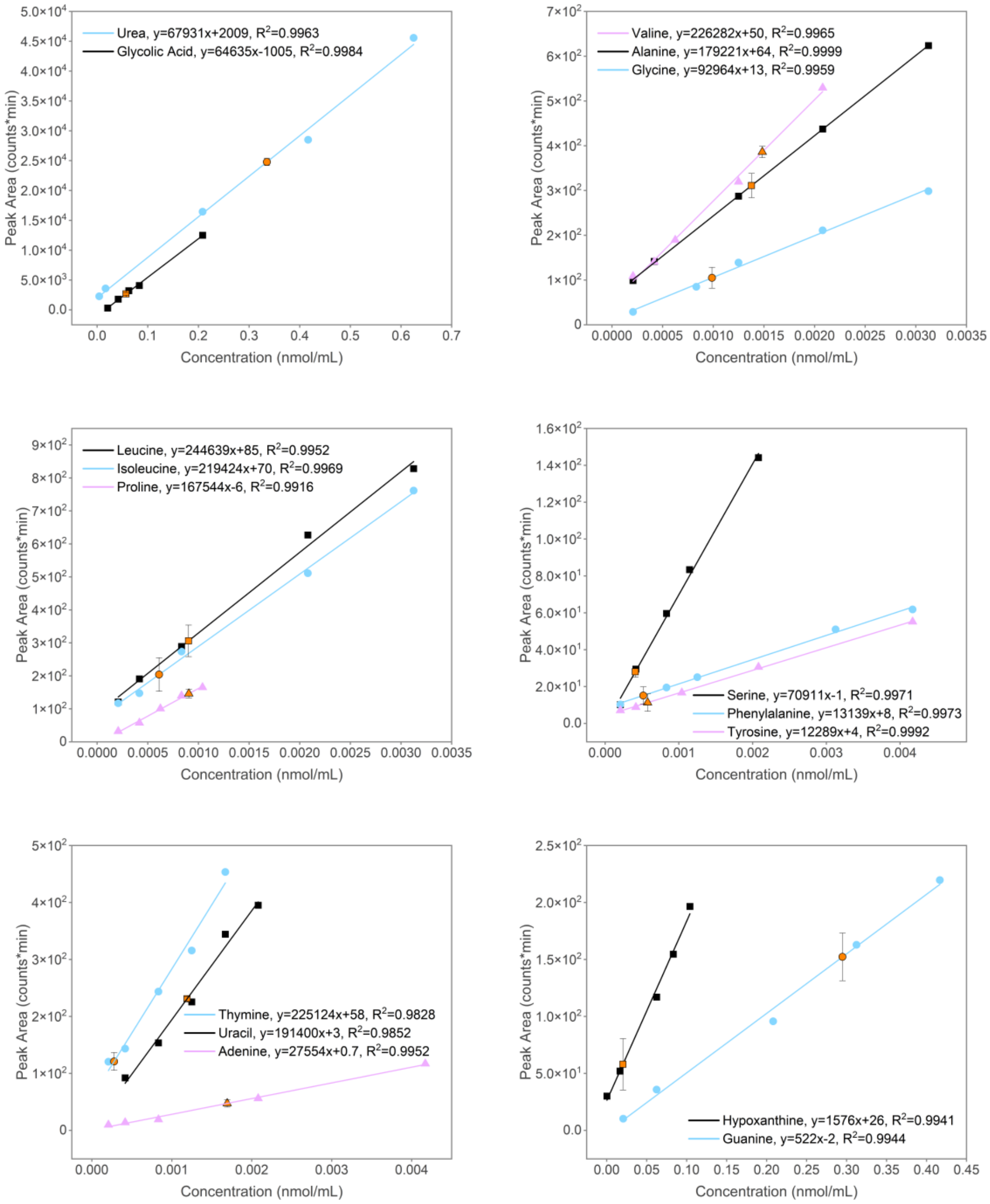

**Figure A3.** Calibration curves for the 9 amino acids, 5 nucleobases, and 2 other biomolecules detected in the Martian analog sample, with orange symbols representing data points obtained from the Martian analog sample.

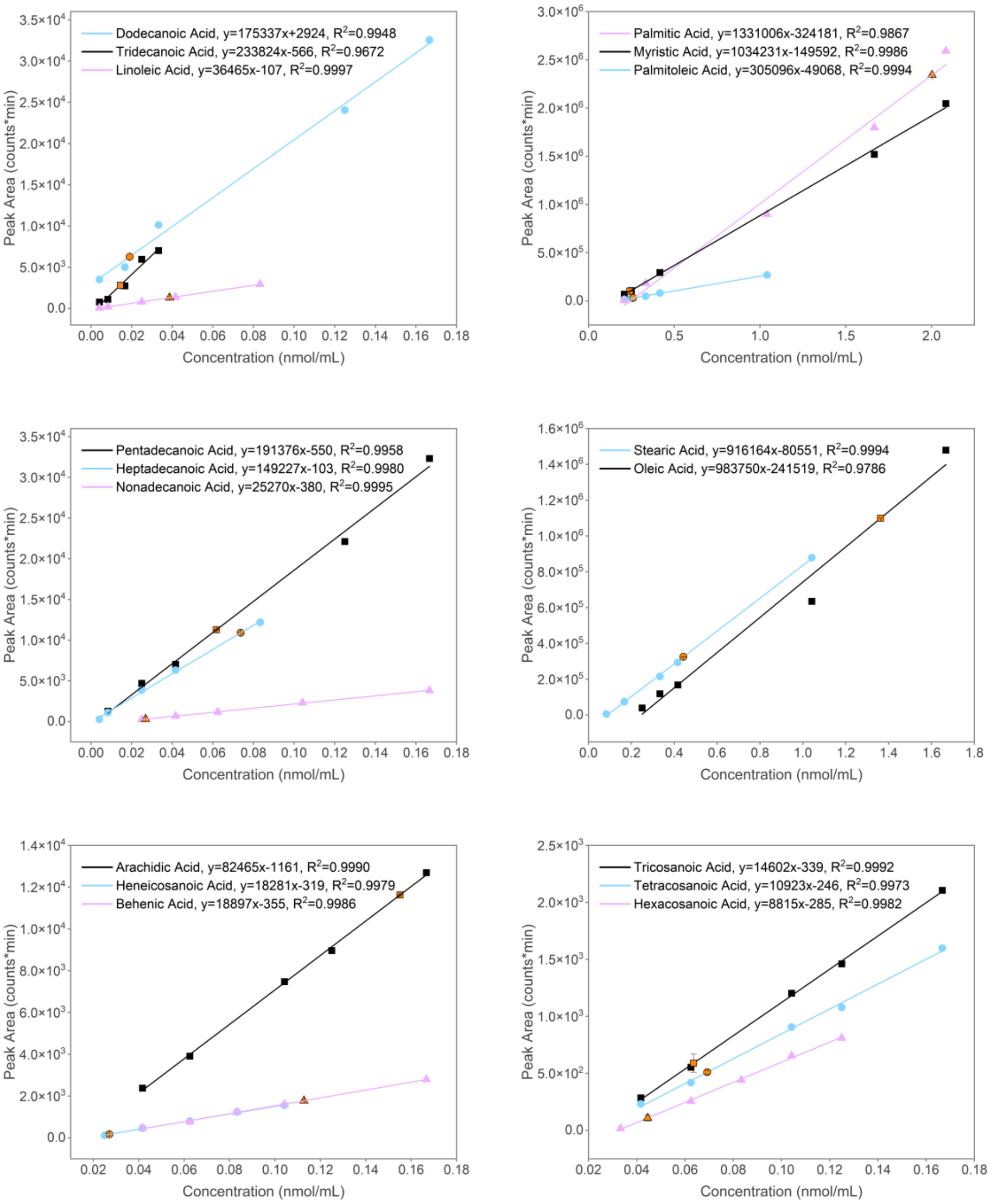

**Figure A4.** Calibration curves for the 17 fatty acids detected in the Martian analog sample, with orange symbols representing data points from the Martian analog sample.

**Table A2.** Concentrations of biomolecules detected in the Titan aerosol analog and the Martian analog sample. The reported values are converted to the concentrations relative to the mass of the original solid analog samples. Quantifier-ion signals meet SNR > 10 for all reported detections, consistent with standard quantitative criteria; two cases with SNR < 10 are indicated. Upper concentration limits for non-detected compounds were estimated from calibration curves, with the limit of detection (LOD) defined as meeting the qualitative criteria (see section 2.2) and requiring at least two ion pairs with SNR > 3. Upper limits are not reported for compounds lacking calibration curves.

| Biomolecule | Titan Aerosol Analog (ppm) | Martian Analog Sample (ppb) |
|---|---|---|
| Amino Acids | | |
| Alanine | 1.4±0.1 | 0.93±0.10 |
| Glycine | 11.6±0.5 | 0.56±0.15 |
| Sarcosine | 1.1±0.2 | <0.14 |
| Valine | <0.17 | 1.3±0.07 |
| Leucine | <0.04 | 0.89±0.21 |
| Isoleucine | <0.04 | 0.60±0.24 |
| Proline | 0.09±0.02 | 0.79±0.08 |
| Methionine | <0.04 | <0.23 |
| Serine | 4.0±0.4 | 0.32±0.04 |
| Threonine | 1.2±0.2 | <0.19 |
| Phenylalanine | 1.3±0.3 | 0.65±0.45 (S/N=8) |
| Aspartic Acid | 1.1±0.4 | <4.2 |
| Hydroxyproline | <0.04 | <0.21 |
| Cysteine | n.d. | n.d. |
| Glutamic Acid | <0.85 | <4.7 |
| Asparagine | n.d. | n.d. |
| Ornithine | n.d. | n.d. |
| Aminoadipic Acid | <1.5 | <8.4 |
| Lysine | n.d. | n.d. |
| Glutamine | n.d. | n.d. |
| Histidine | n.d. | n.d. |
| Tyrosine | 2.2±0.6 | 0.79±0.39 (S/N=7) |
| Tryptophan | n.d. | n.d. |
| Cystine | <14 | <75 |
| Nucleobases | | |
| Uracil | 79.8±6.7 | 1.0±0.11 |
| Thymine | 7.4±1.4 | 0.27±0.10 |
| Cytosine | 161±6.5 | <0.17 |
| Hypoxanthine | 2110±207 | 21±15 |
| Adenine | 1006±39 | 1.7±0.3 |
| Xanthine | 2241±112 | <4.8 |
| Guanine | 7437±352 | 336±49 |
| Fatty Acids | | |
| Dodecanoic Acid | 8.7±0.6 | 26.6±5.7 |
| Tridecanoic Acid | 1.3±0.1 | 21.6±4.0 |
| Myristic Acid | 26.6±1.5 | 384±43 |
| Pentadecanoic Acid | 12.9±1.3 | 104±7.2 |
| Palmitoleic Acid | 129±19 | 461±15 |
| Palmitic Acid | 298±18 | 3572±285 |
| Heptadecanoic Acid | 12.8±1.7 | 138±4.3 |
| Linoleic Acid | 17.2±2.1 | 77±3.0 |
| Oleic Acid | 36.3±3.6 | 2677±276 |
| Stearic acid | 240±14 | 876±19 |
| Nonadecanoic acid | 36.8±1.1 | 55.9±3.6 |
| Arachidic acid | 41.3±1.0 | 337±7.5 |
| Heneicosanoic acid | 8.5±1.8 | 61.6±4.8 |
| Behenic acid | 10.9±2.0 | 267±8.7 |
| Tricosanoic acid | 9.8±2.1 | 157±14 |
| Tetracosanoic acid | 33.5±2.8 | 177±11 |
| Pentacosanoic acid | 12.4±3.5 | <2.2 |
| Hexacosanoic acid | 108±2.3 | 123±11 |
| Other Biomolecules | | |
| Glycolic Acid | 6.9±0.5 | 32.7±2.6 |
| Urea | 1418±314 | 151.7±8.6 |
| Guanidine | 35471±2290 | n.d. |
| Cystathioine | <10 | <56 |